\documentclass[preprint,preprintnumbers,amssymb,prb,superscriptaddress]{revtex4}
\usepackage{graphicx}

\begin{document}

\title{Chiral tunneling and the Klein paradox in graphene}
\author{M. I. Katsnelson}
\affiliation{Institute for Molecules and Materials, Radboud
University Nijmegen, 6525 ED Nijmegen, The Netherlands}
\author{K. S. Novoselov}
\affiliation{Manchester Centre for Mesoscience and
Nanotechnology, University of Manchester, Manchester M13 9PL, UK}
\author{A. K. Geim}
\affiliation{Manchester Centre for Mesoscience and Nanotechnology,
University of Manchester, Manchester M13 9PL, UK}

\begin{abstract}
{\bf The so-called Klein paradox - unimpeded penetration of
relativistic particles through high and wide potential barriers - is
one of the most exotic and counterintuitive consequences of quantum
electrodynamics (QED). The phenomenon is discussed in many contexts
in particle, nuclear and astro- physics but direct tests of the
Klein paradox using elementary particles have so far proved
impossible. Here we show that the effect can be tested in a
conceptually simple condensed-matter experiment by using
electrostatic barriers in single- and bi-layer graphene. Due to the
chiral nature of their quasiparticles, quantum tunneling in these
materials becomes highly anisotropic, qualitatively different from
the case of normal, nonrelativistic electrons. Massless Dirac
fermions in graphene allow a close realization of Klein's gedanken
experiment whereas massive chiral fermions in bilayer graphene offer
an interesting complementary system that elucidates the basic
physics involved.}
\end{abstract}

\maketitle

The term Klein paradox
\cite{klein,greiner,su,grib,dombey,dombey1,krekora} usually refers
to a counterintuitive relativistic process in which an incoming
electron starts penetrating through a potential barrier if its
height $V_0$ exceeds twice the electron's rest energy $mc^2$ (where
$m$ is the electron mass and $c$ the speed of light). In this case,
the transmission probability $T$ depends only weakly on the barrier
height, approaching the perfect transparency for very high barriers,
in stark contrast to the conventional, nonrelativistic tunneling
where $T$ exponentially decays with increasing $V_0$. This
relativistic effect can be attributed to the fact that a
sufficiently strong potential, being repulsive for electrons, is
attractive for positrons and results in positron states inside the
barrier, which align in energy with the electron continuum outside
\cite{su,dombey,dombey1}. Matching between electron and positron
wavefunctions across the barrier leads to the high-probability
tunneling described by the Klein paradox \cite{krekora}. The
essential feature of QED responsible for the effect is the fact that
states at positive and negative energies (electrons and positrons)
are intimately linked (conjugated), being described by different
components of the same spinor wavefunction. This fundamental
property of the Dirac equation is often referred to as the
charge-conjugation symmetry. Although Klein's gedanken experiment is
now well understood, the notion of paradox is still used widely
\cite{greiner,su,grib,dombey,dombey1,krekora}, perhaps because the
effect has never been observed experimentally. Indeed, its
observation requires a potential drop $\approx mc^2$ over the
Compton length $\hbar/mc$, which yields enormous electric
fields~\cite{greiner,grib}(${\cal E} > 10^{16} V/cm$) and makes the
effect relevant only for such exotic situations as, for example,
positron production around super-heavy nuclei~\cite{greiner,grib}
with charge $Z \geq 170$  or evaporation of black holes through
generation of particle-antiparticle pairs near the event horizon
\cite{page}. The purpose of this paper is to show that graphene - a
recently found allotrope of carbon \cite{kostya1} - provides an
effective medium ("vacuum") where relativistic quantum tunneling
described by the Klein paradox and other relevant QED phenomena
could be tested experimentally.

DIRAC-LIKE QUASIPARTICLES IN GRAPHENE

Graphene is a single layer of carbon atoms densely packed in a
honeycomb lattice, or it can be viewed as an individual atomic plane
pulled out of bulk graphite. From the point of view of its
electronic properties, graphene is a two-dimensional zero-gap
semiconductor with the energy spectrum shown in Fig. 1a and its
low-energy quasiparticles are formally described by the Dirac-like
Hamiltonian~\cite{slon,semenoff,haldane}
\begin{equation}
\widehat{H}_{0}=-i\hbar v_{F}\mathbf{\sigma }\nabla \label{zero}
\end{equation}
where $v_{F}$ $\approx10^{6}$ ms$^{-1}$ is the Fermi velocity, and
$\mathbf{\sigma=}\left( \sigma_{x},\sigma _{y}\right)$ are the Pauli
matrices. Neglecting many-body effects, this description is accurate
theoretically\cite{slon,semenoff,haldane} and has also been proven
experimentally\cite{kostya2,kim} by measuring the energy-dependent
cyclotron mass in graphene (which yields its linear energy spectrum)
and, most clearly, by the observation of a relativistic analogue of
the integer quantum Hall effect.

\begin{figure}[t]
\begin{center}\leavevmode
\includegraphics[width=0.39\linewidth]{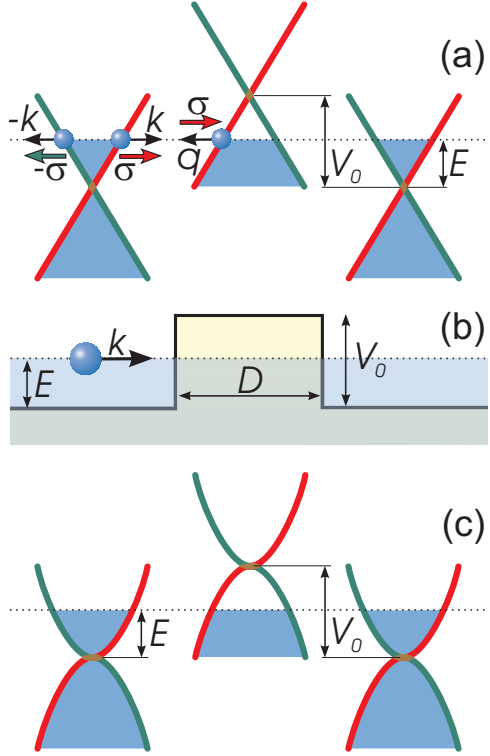}
\caption{Tunneling through a potential barrier in graphene. (a)-
Schematic diagrams of the spectrum of quasiparticles in single-layer
graphene. The spectrum is linear at low Fermi energies ($<$1 eV).
The red and green curves emphasize the origin of the linear
spectrum, which is the crossing between the energy bands associated
with crystal sublattices A and B. The three diagrams illustrate
schematically the positions of the Fermi energy $E$ across the
potential barrier of height $V_0$ and width $D$ shown in (b). The
Fermi level (dotted lines) lies in the conduction band outside the
barrier and the valence band inside it. The blue filling indicates
occupied states. The pseudospin denoted by vector $\sigma$ is
parallel (antiparallel) to the direction of motion of electrons
(holes), which also means that $\sigma$ keeps a fixed direction
along the red and green branches of the electronic spectrum. (c) -
Low-energy spectrum for quasiparticles in bilayer graphene. The
spectrum is isotropic and, despite its parabolicity, also originates
from the intersection of energy bands formed by equivalent
sublattices, which ensures charge conjugation, similar to the case
of single-layer graphene.} \label{Barrier}
\end{center}
\end{figure}
The fact that charge carriers in graphene are described by the
Dirac-like equation (1) rather than the usual Schr\"{o}dinger
equation can be seen as a consequence of graphene's crystal
structure, which consists of two equivalent carbon sublattices A and
B~\cite{slon,semenoff,haldane}. Quantum mechanical hopping between
the sublattices leads to the formation of two cosine-like energy
bands, and their intersection near the edges of the Brillouin zone
(shown in red and green in Fig.~\ref{Barrier}a) yields the conical
energy spectrum. As a result, quasiparticles in graphene exhibit the
linear dispersion relation $E = \hbar k v_F$, as if they were
massless relativistic particles (for example, photons) but the role
of the speed of light is played here by the Fermi velocity
$v_F\approx c/300$. Due to the linear spectrum, one can expect that
graphene's quasiparticles behave differently from those in
conventional metals and semiconductors where the energy spectrum can
be approximated by a parabolic (free-electron-like) dispersion
relation.

Although the linear spectrum is important, it is not the only
essential feature that underpins the description of quantum
transport in graphene by the Dirac equation. Above zero energy, the
current carrying states in graphene are, as usual, electron-like and
negatively charged. At negative energies, if the valence band is not
full, its unoccupied electronic states behave as positively charged
quasiparticles (holes), which are often viewed as a condensed-matter
equivalent of positrons. Note however that electrons and holes in
condensed matter physics are normally described by separate
Schr\"{o}dinger equations, which are not in any way connected (as a
consequence of the Seitz sum rule \cite{VK}, the equations should
also involve different effective masses). In contrast, electron and
hole states in graphene are interconnected, exhibiting properties
analogous to the charge-conjugation symmetry in
QED\cite{slon,semenoff,haldane}. For the case of graphene, the
latter symmetry is a consequence of its crystal symmetry because
graphene's quasiparticles have to be described by two-component
wavefunctions, which is needed to define relative contributions of
sublattices A and B in quasiparticles' make-up. The two-component
description for graphene is very similar to the one by spinor
wavefunctions in QED but the 'spin' index for graphene indicates
sublattices rather than the real spin of electrons and is usually
referred to as pseudospin $\sigma$.

There are further analogies with QED. The conical spectrum of
graphene is the result of intersection of the energy bands
originating from sublattices A and B (see Fig. 1a) and, accordingly,
an electron with energy $E$ propagating in the positive direction
originates from the same branch of the electronic spectrum (shown in
red) as the hole with energy $-E$ propagating in the opposite
direction. This yields that electrons and holes belonging to the
same branch have pseudospin $\sigma$ pointing in the same direction,
which is parallel to the momentum for electrons and antiparallel for
holes (see Fig. 1a). This allows one to introduce chirality
\cite{haldane}, that is formally a projection of pseudospin on the
direction of motion, which is positive and negative for electrons
and holes, respectively. The term chirality is often used to refer
to the additional built-in symmetry between electron and hole parts
of graphene's spectrum (as indicated by color in Fig. 1) and is
analogous (although not completely identical \cite{semenoff,npb}) to
the chirality in three-dimensional QED.

KLEIN PARADOX REFORMULATED FOR SINGLE-LAYER GRAPHENE

Because quasiparticles in graphene accurately mimic Dirac
fermions in QED, this condensed matter system makes it possible to
set up a tunneling experiment similar to that analyzed by Klein. The
general scheme of such an experiment is shown in Fig. 1, where we
consider the potential barrier that has a rectangular shape and is
infinite along the y-axis:
\begin{equation}
V\left( x\right) =\left\{
\begin{array}{cc}
V_{0}, & 0<x<D, \\
0 & \text{otherwise.}
\end{array}
\right.   \label{bar}
\end{equation}
This local potential barrier inverts charge carriers underneath it,
creating holes playing the role of positrons, or vice versa. For
simplicity, we assume in (2) infinitely sharp edges, which allows a
direct link to the case usually considered in
QED~\cite{klein,greiner,su,grib,dombey,dombey1,krekora}. The
sharp-edge assumption is justified if the Fermi wavelength $\lambda$
of quasiparticles is much larger than the characteristic width of
the edge smearing, which in turn should be larger than the lattice
constant (to disallow Umklapp scattering between different valleys
in graphene)~\cite{ando}. Such a barrier can be created by the
electric field effect using a thin insulator or by local chemical
doping~\cite{kostya1,kostya2,kim}. Importantly, Dirac fermions in
graphene are massless and, therefore, there is no formal theoretical
requirement for the minimal electric field ${\cal E}$ to form
positron-like states under the barrier. To create a well-defined
barrier in realistic graphene samples with a disorder, fields ${\cal
E} \approx 10^{5} V/cm$ routinely used in
experiments~\cite{kostya1,kim} should be sufficient, which is eleven
orders of magnitude lower than the fields necessary for the
observation of the Klein paradox for elementary particles.

It is straightforward to solve the tunneling problem sketched in
Fig. 1b. We assume that the incident electron wave propagates at an
angle $\phi$ with respect to the $x$ axis and then try the
components of the Dirac spinor $\psi_1$ and $\psi_2$ for the
Hamiltonian $H=H_{0}+ V\left(x\right)$ in the following form:
\begin{eqnarray}
\psi _{1}\left( x,y\right)  &=&\left\{
\begin{array}{cc}
\left( e^{ik_{x}x}+re^{-ik_{x}x}\right) e^{ik_{y}y}, & x<0, \\
\left( ae^{iq_{x}x}+be^{-iq_{x}x}\right) e^{ik_{y}y}, & 0<x<D, \\
te^{ik_{x}x+ik_{y}y}, & x>D,
\end{array}
\right.   \nonumber \\
\psi _{2}\left( x,y\right)  &=& \left\{
\begin{array}{cc}
s\left( e^{ik_{x}x+i\phi }-re^{-ik_{x}x-i\phi }\right) e^{ik_{y}y},
& x<0,
\\
s^{\prime }\left( ae^{iq_{x}x+i\theta }-be^{-iq_{x}x-i\theta
}\right)
e^{ik_{y}y}, & 0<x<D, \\
ste^{ik_{x}x+ik_{y}y+i\phi }, & x>D,
\end{array}
\right.
\end{eqnarray}
where $k_{F} = 2\pi/\lambda$ is the Fermi wavevector,
$k_{x}=k_{F}\cos \phi$ and $k_{y}=k_{F}\sin \phi$ are the wavevector
components outside the barrier, $q_{x}=\sqrt{\left( E-V_{0}\right)
^{2}/\hbar^{2}v_{F}^{2}-k_{y}^{2}},$ $\theta =\tan
^{-1}\left(k_{y}/q_{x}\right) $ is the refraction angle, $s=signE$,
 $s^{\prime}=sign\left( E-V_{0}\right)$. Requiring the continuity
of the wavefunction by matching up coefficients $a, b, t, r$, we
find the following expression for the reflection coefficient $r$
\begin{equation}
r=2ie^{i\phi }\sin \left( q_{x}D\right) \frac{\sin \phi -ss^{\prime
}\sin \theta }{ss^{\prime }\left[ e^{-iq_{x}D}\cos \left( \phi
+\theta \right) +e^{iq_{x}D}\cos \left( \phi -\theta \right) \right]
-2i\sin \left( q_{x}D\right) }. \label{R1}
\end{equation}

\begin{figure}[t]
\begin{center}\leavevmode
\includegraphics[width=0.4\linewidth]{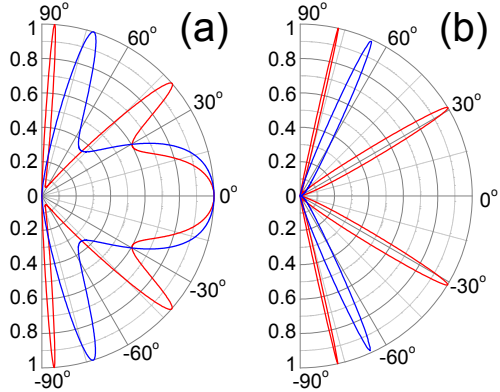}
\caption{Klein-like quantum tunneling in graphene systems.
Transmission probability $T$ through a 100-nm-wide barrier as a
function of the incident angle for (a) single- and (b) bi-layer
graphene. The electron concentration $n$ outside the barrier is
chosen $0.5\times 10^{12}$ cm$^{-2}$ for all cases. Inside the
barrier, hole concentrations $p$ are $1\times10^{12}$ and
$3\times10^{12}$ cm$^{-2}$ for red and blue curves, respectively
(such concentrations are most typical in experiments with graphene).
This corresponds to the Fermi energy $E$ of incident electrons
$\approx 80$ and $17$ meV for single- and bi-layer graphene,
respectively, and $\lambda\approx 50$ nm. The barrier heights
$V_{0}$ are (a) 200 and (b) 50 meV (red curves) and (a) 285 and (b)
100 meV (blue curves).} \label{Angular}
\end{center}
\end{figure}

Fig.~\ref{Angular}a shows examples of the angular dependence of
transmission probability $T = \left|
t\right|^{2}=1-\left|r\right|^{2}$ calculated using the above
expression. In the limit of high barriers $\left| V_{0}\right| \gg
\left| E\right|$, the expression for $T$ can be simplified to
\begin{equation}
T=\frac{\cos ^{2}\phi }{1-\cos ^{2}\left( q_{x}D\right) \sin^{2}\phi
}.
\end{equation}
Equations (4,5) yield that under resonance conditions $q_{x}D= \pi
N$ , $N=0,\pm 1, \ldots$ the barrier becomes transparent ($T=1$).
More significantly, however, the barrier remains \textit{always}
perfectly transparent for angles close to the normal incidence $\phi
=0$. The latter is the feature unique to massless Dirac fermions and
directly related to the Klein paradox in QED. One can understand
this perfect tunneling in terms of the conservation of pseudospin.
Indeed, in the absence of pseudospin-flip processes (such processes
are rare as they require a short-range potential, which would act
differently on A and B sites of the graphene lattice), an electron
moving to the right can be scattered only to a right-moving electron
state or left-moving hole state. This is illustrated in Fig. 1a,
where charge carriers from the "red" branch of the band diagram can
be scattered into states within the same "red" branch but cannot be
transformed into any state on the "green" branch. The latter
scattering event would require the pseudospin to be flipped. The
matching between directions of pseudospin $\sigma$ for
quasiparticles inside and outside the barrier results in perfect
tunneling. In the strictly one-dimensional case, such perfect
transmission of Dirac fermions has been discussed in the context of
electron transport in carbon nanotubes~\cite{ando,mceuen} (see also
ref. [19]). Our analysis extends this tunneling problem to the
two-dimensional (2D) case of graphene.

CHIRAL TUNNELING IN BILAYER GRAPHENE

To elucidate which features of the anomalous tunneling in graphene
are related to the linear dispersion and which to the pseudospin and
chirality of the Dirac spectrum, it is instructive to consider the
same problem for bilayer graphene. There are both differences and
similarities between the two graphene systems. Indeed, charge
carriers in bilayer graphene have parabolic energy spectrum as shown
in Fig. 1c, which means they are massive quasiparticles with a
finite density of states at zero energy, similar to conventional
nonrelativistic electrons. On the other hand, these quasiparticles
are also chiral and described by spinor wavefunctions
\cite{bilayer,falko}, similar to relativistic particles or
quasiparticles in single-layer graphene. Again, the origin of the
unusual energy spectrum can be traced to the crystal lattice of
bilayer graphene with four equivalent sublattices \cite{falko}.
Although ``massive chiral fermions'' do not exist in the field
theory their existence in the condensed matter physics (confirmed
experimentally \cite{bilayer}) offers a unique opportunity to
clarify the importance of chirality in the relativistic tunneling
problem described by the Klein paradox. In addition, the relevant
QED-like effects appear to be more pronounced in bilayer graphene
and easier to test experimentally, as discussed below.

Charge carriers in bilayer graphene are described by an off-diagonal
Hamiltonian~\cite{bilayer,falko}
\begin{equation}
\widehat{H}_{0}=-\frac{\hbar^{2}}{2m}\left(
\begin{array}{cc}
0 & \left( k_{x}-ik_{y}\right)^{2} \\
 \left( k_{x}+ik_{y}\right)^{2} & 0
\end{array}
\right) \label{hamiltonianbilayer}
\end{equation}
which yields a gapless semiconductor with chiral electrons and holes
having a finite mass $m$. An important formal difference between the
tunneling problems for single- and bi- layer graphene is that in the
latter case there are \textit{four} possible solutions for a given
energy $E=\pm \hbar^{2}k_{F}^{2}/2m$. Two of them correspond to
propagating waves and the other two to evanescent ones. Accordingly,
for constant potential $V_{i}$, eigenstates of
Hamiltonian~(\ref{hamiltonianbilayer}) should be written as
\begin{eqnarray}
\psi _{1}\left( x,y\right)  &=&\left(
a_{i}e^{ik_{ix}x}+b_{i}e^{-ik_{ix}x}+c_{i}e^{\kappa
_{ix}x}+d_{i}e^{-\kappa
_{ix}x}\right) e^{ik_{y}x}  \nonumber \\
\psi _{2}\left( x,y\right)  &=&s_{i}\left( a_{i}e^{ik_{ix}x+2i\phi
_{i}}+b_{i}e^{-ik_{ix}x-2i\phi _{i}}-c_{i}h_{i}e^{\kappa _{ix}x}-\frac{d_{i}%
}{h_{i}}e^{-\kappa _{ix}x}\right) e^{ik_{y}y} \label{wavebilayer}
\end{eqnarray}
where
\begin{displaymath}
s_{i}=sign\left(V_{i}-E\right); \hspace{0.5cm} \hbar
k_{ix}=\sqrt{2m\left| E-V_{i}\right| }\cos \phi _{i}; \hspace{0.5cm}
\hbar k_{iy}=\sqrt{2m\left| E-V_{i}\right|}\sin \phi _{i} $$$$
\kappa _{ix}=\sqrt{k_{ix}^{2}+2k_{iy}^{2}}; \hspace{1cm}
h_{i}=\left( \sqrt{1+\sin ^{2}\phi _{i}}-\sin \phi _{i}\right) ^{2}.
\end{displaymath}

To find the transmission coefficient through barrier (\ref{bar}),
one should set $d_{1}=0$ for $x<0,$ $b_{3}=c_{3}=0$ for $x>D$ and
satisfy the continuity conditions for both components of the
wavefunction and their derivatives. For the case of an electron beam
that is incident normally ($\phi =0$) and low barriers $V_{0} < E$
(over-barrier transmission), we obtain $\psi_1 = - \psi_2$ both
outside and inside the barrier, and the chirality of fermions in
bilayer graphene does not manifest itself. In this case, scattering
at the barrier (\ref{bar}) is the same as for electrons described by
the Schr\"{o}dinger equation. However, for any finite $\phi$ (even
in the case $V_{0} < E$), waves localized at the barrier interfaces
are essential to satisfy the boundary conditions.

\begin{figure}[t]
\begin{center}\leavevmode
\includegraphics[width=0.4\linewidth]{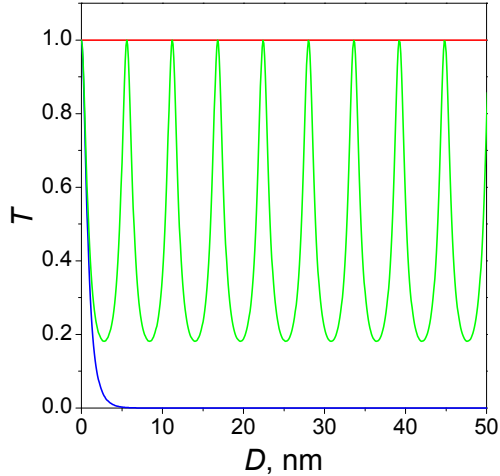}
\caption{Transmission probability $T$ for normally incident
electrons in single-  and bi- layer graphene (red and blue curves,
respectively) and in a non-chiral zero-gap semiconductor (green
curve) as a function of width $D$ of the tunnel barrier.
Concentrations of charge carriers are chosen as $n=0.5\times
10^{12}$ cm$^{-2}$ and $p=1\times 10^{13}$ cm$^{-2}$ outside and
inside the barrier, respectively, for all three cases. This yields
barrier's height of $\sim450$ meV for  graphene and $\sim240$ meV
for the other two materials. Note that the transmission probability
for bilayer graphene decays exponentially with the barrier width,
even though there are plenty of electronic states inside the
barrier.} \label{BarrierWidth}
\end{center}
\end{figure}

The most intriguing behavior is found for $V_{0}>E$, where electrons
outside the barrier transform into holes inside it, or vice versa.
Examples of the angular dependence of $T$ in bilayer graphene are
plotted in Fig.~\ref{Angular}b. They show a dramatic difference as
compared with the case of massless Dirac fermions. There are again
pronounced transmission resonances at some incident angles, where
$T$ approaches unity. However, instead of the perfect transmission
found for normally-incident Dirac fermions (see
Fig.~\ref{Angular}a), our numerical analysis has yielded the
opposite effect: Massive chiral fermions are always perfectly
reflected for angles close to $\phi =0$.

Accordingly, we have analyzed this case in more detail and found the
following analytical solution for the transmission coefficient $t$:
\begin{equation}
t=\frac{4ik_{1}k_{2}}{\left( k_{2}+ ik_{1}\right)^{2}
e^{-k_{2}D}-\left( k_{2}- ik_{1}\right)^{2} e^{k_{2}D}},
\label{incident}
\end{equation}
where subscripts 1,2 label the regions outside and inside the
barrier, respectively. Particularly interesting is the case of a
potential step, which corresponds to a single $p-n$ junction. Eq
(\ref{incident}) shows that such a junction should completely
reflect a normally-incident beam ($T=0$). This is highly unusual
because the continuum of electronic states at the other side of the
step is normally expected to allow some tunneling. Furthermore, for
a single $p-n$ junction with $V_{0}\gg E$, the following analytical
solution for any $\phi$ has been found:
\begin {equation}
T= \frac{E}{V_{0}}sin^2(2\phi)
\end{equation}
which again yields $T =0$ for $\phi =0$. This behavior is in obvious
contrast to single-layer graphene, where normally-incident electrons
are always perfectly transmitted.

The perfect reflection (instead of the perfect transmission) can be
viewed as another incarnation of the Klein paradox, because the
effect is again due to the charge-conjugation symmetry (fermions in
single- and bi-layer graphene exhibit chiralities that resemble
those associated with spin 1/2 and 1, respectively)
\cite{bilayer,falko}. For single-layer graphene, an electron
wavefunction at the barrier interface matches perfectly the
corresponding wavefunction for a hole with the same direction of
pseudospin (see Fig. 1a), yielding $T=1$. In contrast, for bilayer
graphene, the charge conjugation requires a propagating electron
with wavevector $k$ to transform into a hole with wavevector $ik$
(rather than $-k$), which is an evanescent wave inside a barrier.

COMPARISON WITH TUNNELING OF NONCHIRAL PARTICLES

For completeness, we compare the obtained results with the case of
normal electrons. If a tunnel barrier contains no electronic states,
the difference is obvious: the transmission probability in this case
is well known to decay exponentially with increasing barrier's width
and height~\cite{esaki} so that the tunnel barriers discussed above
would reflect electrons completely. However, both graphene systems
are gapless, and it is more appropriate to compare them with gapless
semiconductors having nonchiral charge carriers (such a situation
can be realized in certain heterostructures~\cite{meyer,teissier}).
In this case, one finds
\begin{equation}
t=\frac{4k_{x}q_{x}}{\left( q_{x}+ k_{x}\right)^{2}
e^{-iq_{x}D}-\left( q_{x}- k_{x}\right)^{2} e^{iq_{x}D}},
\label{normal}
\end{equation}
where $k_{x}$ and $q_{x}$ are $x$-components of the wave vector
outside and inside the barrier, respectively. Again, similarly to
the case of single- and bi-layer graphene, there are resonance
conditions $q_{x}D= \pi N,$ $N=0,\pm 1,...$ at which the barrier is
transparent. For the case of normal incidence ($\phi=0$) the
tunneling coefficient is then an oscillating function of tunneling
parameters and can exhibit any value from 0 to 1 (see Fig. 3). This
is in contrast to graphene, where $T$ is always 1, and bilayer
graphene, where $T=0$ for sufficiently wide barriers $D
> \lambda$. This makes it clear that the drastic difference between
the three cases is essentially due to different chiralities or
pseudospins of the quasiparticles involved rather than any other
feature of their energy spectra.

\begin{figure}[t]
\begin{center}\leavevmode
\includegraphics[width=0.4\linewidth]{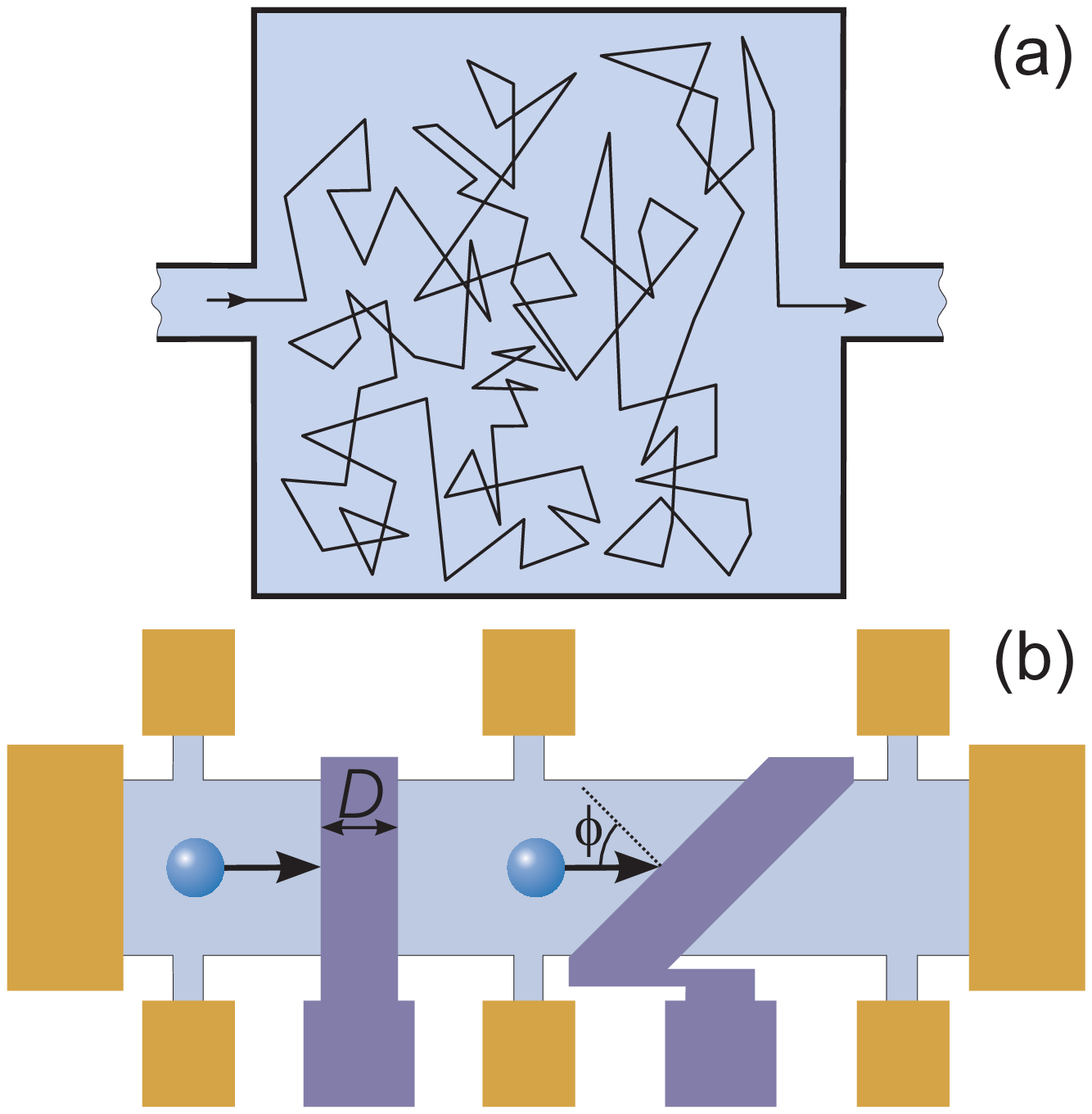}
\caption{The chiral nature of quasiparticles in graphene strongly
affects its transport properties. (a) A diffusive conductor of a
size smaller than the phase-coherence length is connected to two
parallel one-dimensional leads. For normal electrons, transmission
probability $T$ through such a system depends strongly on the
distribution of scatterers. In contrast, for massless Dirac
fermions, $T$ is always equal to unity due to the additional memory
about the initial direction of pseudospin (see text). (b) Schematic
diagram of one of the possible tunneling experiments in graphene.
Graphene (light-blue) has two local gates (dark-blue) that create
potential barriers of a variable height. The voltage drop across the
barriers is measured by using potential contacts shown in orange.}
\label{Experiment}
\end{center}
\end{figure}

IMPLICATIONS FOR EXPERIMENT

The found tunneling anomalies in the two graphene systems are
expected to play an important role in their transport properties,
especially in the regime of low carrier concentrations where
disorder induces significant potential barriers and the systems are
likely to split into a random distribution of p-n junctions. In
conventional 2D systems, strong enough disorder results in
electronic states that are separated by barriers with exponentially
small transparency~\cite{ziman,lifshitz}. This is known to lead to
the Anderson localization. In contrast, in both graphene materials
all potential barriers are relatively transparent ($T\approx1$ at
least for some angles) which does not allow charge carriers to be
confined by potential barriers that are smooth on atomic scale.
Therefore, different electron and hole ``puddles'' induced by
disorder are not isolated but effectively percolate, thereby
suppressing localization. This consideration can be important for
the understanding of the minimal conductivity $\approx e^2/h$
observed experimentally in both single-layer~\cite{kostya2} and
bilayer~\cite{bilayer} graphene.

To elucidate further the dramatic difference between quantum
transport of Dirac fermions in graphene and normal 2D electrons,
Fig.~\ref{Experiment}a suggests a gedanken experiment where a
diffusive conductor is attached to ballistic one-dimensional leads,
as in the Landauer formalism. For conventional 2D systems,
transmission and reflection coefficients through such a conductor
are sensitive to detailed distribution of impurities and a shift
of a single impurity by a distance of the order of $\lambda$ can
completely change the coefficients \cite{meso}. In contrast, the
conservation of pseudospin in graphene strictly forbids
backscattering and makes the disordered region in
Fig.~\ref{Experiment}a \textit{always} completely transparent,
independent of disorder (as long as it is smooth on the scale of the
lattice constant~\cite{ando}). This extension of the Klein problem
to the case of a random scalar potential has been proven by using
the Lippmann-Schwinger equation (see the Supplementary Information).
Unfortunately, this particular experiment is probably impossible to
realize in practice because scattering at graphene's edges does not
conserve the pseudospin~\cite{ando,berry}. Nevertheless, the above
consideration shows that impurity scattering in the bulk of graphene
should be suppressed as compared to the normal conductors.

The above analysis shows that the Klein paradox and associated
relativistic-like phenomena can be tested experimentally using
graphene devices. The basic principle behind such experiments would
be to employ local gates and collimators similar to those used in
electron optics in 2D gases~\cite{optics1, optics2}. One possible
experimental setup is shown schematically in Fig.~\ref{Experiment}b.
Here, local gates simply cross the whole graphene sample at
different angles (for example, $90^o$ and $45^o$). Intrinsic
concentrations of charge carriers are usually low ($\sim10^{11}$
cm$^{-2}$), whereas concentrations up to $1\times10^{13}$ cm$^{-2}$
can be induced under the gated regions by the bipolar electric field
effect~\cite{kostya1}. This allows potential barriers with heights
up to  $V_0\approx0.4$ eV and $\approx0.23$ eV for single- and
double-layer samples, respectively. By measuring the voltage drop
across the barriers as a function of applied gate voltage, one can
analyze their transparency for different $V_0$. Our results in
Fig.~\ref{Angular} show that for graphene the $90^o$ barrier should
exhibit low resistance and no significant changes in it with
changing gate voltage. In comparison, the $45^o$ barrier is
expected to have much higher resistance and show a number of
tunneling resonances as a function of gate voltage. The situation
should be qualitatively different for bilayer graphene, where local
barriers should result in a high resistance for the perpendicular
barrier and pronounced resonances for the $45^o$ barrier.

Furthermore, the fact that a barrier (or even a single $p-n$
junction) incorporated in a bilayer graphene device should lead to
exponentially small tunneling current can be exploited in developing
graphene-based field effect transistors (FET). Such transistors are
particularly tempting because of their high mobility and ballistic
transport at submicron distances \cite{kostya1,kostya2,kim}.
However, the fundamental problem along this route is that the
conducting channel in single-layer graphene cannot be pinched off
(because of the minimal conductivity), which severely limits
achievable on-off ratios for such FETs \cite{kostya1} and,
therefore, the scope for their applications. A bilayer FET with a
local gate inverting the sign of charge carriers should yield much
higher on-off ratios.

OUTLOOK

We have shown that the recently found two carbon allotropes provide
an effective medium for mimicking relativistic quantum effects. On
the one hand, this allows one to set up such exotic experiments as
the one described by the Klein paradox and could be useful for
analysis of other relevant QED problems. On the other hand, our work
also shows that the known QED problems and their solutions can be
applied to graphene to achieve better understanding of transport
properties of this unique material that is interesting from the view
point of both fundamental physics and applications.

\textit{Acknowledgements}. We are grateful to Antonio Castro Neto,
Vladimir Fal'ko, Paco Guinea and Dmitri Khveshchenko for
illuminating discussions. This work was supported by EPSRC (UK) and
FOM (Netherlands).

\end{document}